\documentclass[%
 aip,
rsi,%
 amsmath,amssymb,
 reprint,%
]{revtex4-1}

\usepackage{graphicx}
\usepackage{dcolumn}
\usepackage{bm}
\usepackage{mathrsfs}
\usepackage{algorithm}
\usepackage{algpseudocode}
\usepackage{xcolor}
\usepackage{soul}
\usepackage[version=4]{mhchem}

\newcommand{\bsym}{\boldsymbol}

\renewcommand{\b}{\bsym}

\newcommand{\ket}[1]{\vert #1 \rangle}

\newcommand{\bra}[1]{\langle #1 \vert}

\newcommand{\Tbraket}[3]{\langle #1 \hspace{.10em} \vert \hspace{.10em} #2 \hspace{.10em} \vert \hspace{.10em} #3 \rangle}

\newcommand{\tbar}{\bar{t}}

\newcommand{\hf}{\mathrm{HF}}

\allowdisplaybreaks
\begin{document}

\title{Efficient implementation of molecular CCSD gradients with Cholesky-decomposed electron repulsion integrals}

\author{Anna Kristina Schnack-Petersen}
\affiliation{ 
Department of Chemistry, Technical University of Denmark, 2800 Kongens Lyngby, Denmark
}
\author{Henrik Koch}
\affiliation{%
Scuola Normale Superiore, Piazza dei Cavaleri 7, 56126 Pisa, Italy 
}%
\affiliation{ 
Department of Chemistry, Norwegian University of Science and Technology, 7491 Trondheim, Norway
}%
\author{Sonia Coriani}
\affiliation{ 
Department of Chemistry, Technical University of Denmark, 2800 Kongens Lyngby, Denmark
}%
\affiliation{ 
Department of Chemistry, Norwegian University of Science and Technology, 7491 Trondheim, Norway
}%
\author{Eirik F.~Kj{\o}nstad}
\email{eirik.kjonstad@ntnu.no}
\affiliation{ 
Department of Chemistry, Norwegian University of Science and Technology, 7491 Trondheim, Norway
}%

\date{\today}

\begin{abstract}
We present an efficient 
implementation of ground and excited state CCSD gradients
based on Cholesky-decomposed electron repulsion integrals.
Cholesky decomposition, like density-fitting,  
is an inner projection method, and thus similar implementation schemes can be applied for both methods.
One well-known advantage of inner projection methods, 
which we exploit in our implementation,
is that 
one can
avoid storing large $V^3O$ and $V^4$ arrays 
by 
instead considering
three-index 
intermediates.
Furthermore,
our 
implementation does not require the formation and storage of Cholesky vector derivatives. 
The new 
implementation is shown to perform well, with 
less than 10\% of the time 
spent
calculating
the gradients in geometry optimizations. 
The computational time spent per optimization cycle are
furthermore
found to be significantly lower compared to other 
implementations based on an inner projection method.
We illustrate the capabilities of the implementation
by optimizing the geometry
of the retinal molecule (\ce{C20H28O}) at the CCSD/aug-cc-pVDZ level of theory.
\end{abstract}

\maketitle

\section{Introduction}
The gradient of the electronic energy with respect to  the nuclear coordinates, known as the molecular gradient,
is a particularly useful quantity in computational chemistry. It is essential for determining both local energy minima and equilibrium geometries\cite{Schlegel2011} and thus 
for predicting the stability and structure of molecular systems, as well as
molecular properties at equilibrium. In addition, the  molecular gradient is 
essential for locating transition state geometries, which can aid in elucidating chemical reaction paths and in estimating reaction rates.\cite{Truhlar2014} Moreover, molecular gradients are also required for
predicting the time evolution of molecular systems, since the gradient provides the forces that act on the atomic nuclei in the absence of sizeable non-adiabatic effects.\cite{Marx2009,Curchod2018}

Over the past 
three decades,
coupled cluster (CC) methods have gained popularity\cite{cizek_66,bartlett_81,bartlett_89,crawford_2000,bartlett_07} due to their high and systematically improvable accuracy. Today, they are widely considered the most efficient 
for describing dynamical correlation 
whenever
the ground state 
is dominated by a single determinant.\cite{bartlett_07, helgaker2014molecular}
Coupled cluster methods that include approximate 
triple excitations
(such as CC3\cite{Koch1997}) are generally considered the state-of-the-art in computational chemistry, 
but they are still  
too costly 
for molecular systems with more than about fifteen second-row elements.\cite{Izak_20} 
Nonetheless, coupled cluster
calculations
are becoming
increasingly feasible, 
particularly 
for methods that include double excitations either approximately, such as CC2,\cite{CHRISTIANSEN1995409} or 
in full, that is, CCSD.\cite{Purvis1982} 
 
It is therefore of
considerable
interest 
to develop efficient implementations of molecular gradients
at the CCSD level of theory, both for the ground and the excited states.
Implementations of such gradients already exist in a number of programs.
Ground state 
gradients are available
 in commercial codes such as Q-Chem\cite{Qchem541,Krylov_19} and Gaussian\cite{g16} and 
 in open-source programs such as Psi4\cite{psi4,Bozkaya_16} and Dalton,\cite{daltonpaper,Hald_02}
as well as in free programs such as CFOUR\cite{cfour, Gauss_91} and MRCC.\cite{MRCC, mihaly_03} 
However, out of the above-mentioned programs,
only Q-Chem, CFOUR, and MRCC 
include excited state gradients at the CCSD level.

Cholesky decomposition (CD) of the electron repulsion integrals\cite{beebe_77,ROEGGEN_86,koch_03,aquilante_07,deprince_13, bozkaya_14} has become a valuable tool for efficient implementations in quantum chemistry. 
Due to the rank-deficiency of the integral matrix, 
CD 
implies significantly reduced computational requirements, both in terms of storage and the number of floating-point operations. The CD method dates back to the 1970s,\cite{beebe_77} but 
it has seen a resurgence of interest over the past decades 
due to improvements in algorithms\cite{koch_03, delcey_14, folkestad_19} and computer hardware. 
These developments have
made 
CD competitive with the prevailing inner projection method, the resolution-of-identity (RI) or density-fitting method.\cite{whitten_73,dunlap_79,FEYEREISEN_93,VAHTRAS93,rendell_94,weigend_02,sodt_06,werner_03,shutz_03,werner_11, deprince_13, bozkaya_14} 
As a result, 
there is currently a demand for
efficient CD-based coupled cluster implementations (e.g., for molecular gradients).
To date, however, such implementations are still rather scarce.
Indeed,
out of the programs mentioned above,
only 
Q-Chem 
offers 
an implementation of CCSD  gradients based on Cholesky-decomposed integrals.\cite{Krylov_19}

The factorized form of the electron repulsion integrals, obtained by inner projection, has an interesting implication for molecular gradient algorithms. Two- and three-index density intermediates naturally arise, allowing one to 
avoid storing
the $V^3 O$ and $V^4$ blocks of the density matrix. 
This 
has been exploited both in 
CD (for CASSCF)\cite{delcey_14} and in RI (for CCSD).\cite{Bozkaya_16} 
The equivalence between CD and RI
means that RI implementations can be adapted to the framework of 
CD, which is the focus of this work.  
An alternative algorithm for CD-CCSD gradients was recently suggested by Feng {\it et al.}\cite{Krylov_19} and implemented in Q-Chem.\cite{Qchem541} However, this algorithm relies on the calculation and storage of Cholesky vector derivatives. 
Such an approach 
is disadvantageous because it
implies
a relatively large O($N^4$) storage requirement, 
thereby
imposing a 
limitation on the 
system size.

In this work we describe
a new and efficient implementation of the ground and excited state CCSD gradients, which we have 
incorporated into
a development version of the open-source program $e^T$ (version 1.4).\cite{etpaper} This implementation is partly based on the one reported by Bozkaya and Sherrill for RI-CCSD,\cite{Bozkaya_16} where the gradient is constructed from two- and three-index density intermediates; for these intermediates, see also the CD-CASSCF implementation by Delcey \emph{et al}.\cite{delcey_14}
Our
implementation is well-suited for large-scale applications
and 
makes use of the recent two-step CD implementation by Folkestad \emph{et al}.\cite{folkestad_19} 
In particular,
our implementation calculates, on-the-fly, derivative integrals involving the auxiliary basis,
ensuring that no O($N^4$) storage requirements are associated with the Cholesky vectors.

\section{Theory}
\subsection{Analytical expression for the molecular gradient}
The molecular CCSD gradient is conveniently derived from the Lagrangian
\begin{align}
\begin{split}
    \mathcal{L} &= E + \sum_{\mu \neq 0}\bar{\zeta}_{\mu}\Omega_{\mu}+\sum_{p\leq q} \bar{\kappa}_{pq}(F_{pq}-\delta_{pq}\epsilon_p) + \bar{\lambda} \mathcal{O} \\
    &=\sum_{pq} h_{pq}D_{pq} + \sum_{pqrs} g_{pqrs}d_{pqrs} 
    + \sum_{\mu \neq 0}\bar{\zeta}_{\mu}\Omega_{\mu} \\
    &\quad+\sum_{p\leq q} \bar{\kappa}_{pq}(F_{pq}-\delta_{pq}\epsilon_p) +  \bar{\lambda} \mathcal{O}.\\
\end{split}
    \label{eq:lagrangian}
\end{align}
In this expression, $E$ is the energy, $h_{pq}$ and $g_{pqrs}$ are the one- and two-electron integrals associated with the Hamiltonian $H$, and $D_{pq}$ and $d_{pqrs}$ are the one- and two-electron densities. Lagrangian multipliers are denoted with a bar ($\bar{\zeta}_{\mu}, \bar{\kappa}_{pq}, \bar{\lambda}$). Furthermore,  
\begin{align}
    \Omega_{\mu} &= \bra{\mu}\bar{H}\ket{\hf},
\end{align}
where $\ket{\hf}$ is the Hartree-Fock state and
\begin{align}
    \bar{H} &= \exp{(-T)}\exp{(\kappa)}H\exp{(-\kappa)}\exp{(T)}.\label{Hbar}
\end{align}
Here $\kappa$ denotes the orbital rotation operator, where it is implied that $\kappa = 0$ at the nuclear geometry where the derivative is to be evaluated.\cite{Helgaker1992,Hald_02} The cluster operator is $T = \sum_\mu t_\mu \tau_\mu$, where $\tau_{\mu}$ is an excitation operator and excited configurations are denoted as $\ket{\mu} = \tau_{\mu} \ket{\hf}$; 
in addition, $\tau_0 = \mathbb{I}$ and hence $\ket{0}=\ket{\hf}$.
Moreover, the Fock matrix is given as
\begin{align}
    F_{pq} = h_{pq} + \sum_k (2 g_{pqkk} - g_{pkkq}) \label{eq:fockmatrix}
\end{align}
and $\epsilon_p$ denotes the energy of the $p$'th molecular orbital (MO). Finally, $\mathcal{O}$ is defined such that $\mathcal{O} = 0$ ensures normalization, i.e., that the left and right coupled cluster states are binormal. This term is described in more detail below.

The integrals in the Fock matrix ($F_{pq}$) are always expressed in the MO basis, whereas the integrals in the coupled cluster energy,
\begin{align}
    E = \sum_{pq} h_{pq}D_{pq} + \sum_{pqrs} g_{pqrs}d_{pqrs},
\end{align}
are either expressed in the MO basis or in the $T_1$-transformed basis. 
Above and throughout, $p$, $q$, $r$, and $s$ are used to denote generic MOs; $i$, $j$, $k$, and $l$ denote occupied MOs; and $a$, $b$, $c$, and $d$ denote virtual MOs. 

The expressions for the energy $E$ and the normalization condition $\mathcal{O}$ depend on whether we are considering the ground state or an excited state. In particular, 
\begin{align}
    E^{\textrm{GS}} &= \bra{\hf}\bar{H}\ket{\hf}\label{Egs}\\
    E^{\textrm{ES}} &= \sum_{\mu\nu \geq 0}L_{\mu}\bra{\mu}\bar{H}\ket{\nu}R_{\nu}\label{Ees}
\end{align}
and
\begin{align}
    \mathcal{O}^\textrm{GS} &= 0 \\
    \mathcal{O}^\textrm{ES} &= 1 - \sum_{\mu \geq 0} L_\mu R_\mu,
\end{align}
where $L_\mu$ and $R_\mu$ denote the left and right excited state amplitudes. For the ground state, normalization is automatically fulfilled,\cite{helgaker2014molecular} and can be ignored as 
$\mathcal{O}^\textrm{GS} = 0$, which effectively removes 
the normalization condition from the Lagrangian.

The stationarity conditions with respect to 
 $\bar{\zeta}_\mu$ and 
 $\bar{\kappa}_{pq}$ are, respectively, the well-known amplitude and canonical Hartree-Fock equations, $\Omega_\mu = 0$ and $F_{pq} = \delta_{pq} \epsilon_p$. 
 Furthermore, the stationarity condition with respect to 
 $\bar{\lambda}$ enforces the biorthonormality constraint.
 The orbital rotation multipliers $\bar{\kappa}_{pq}$ are  determined from the stationarity condition
{with respect to the orbital rotation parameters $\kappa_{pq}$,~\cite{CI:derivatives,Helgaker1992,Hald_02}}
\begin{align}
    \dfrac{\partial \mathcal{L}}{\partial \kappa_{pq}} =0 \iff \boldsymbol{\bar{\kappa}}~
    {}^{\boldsymbol{\bar{\kappa}}}\!{\boldsymbol{A}} =-~\bm{^{\bar{\kappa}}\!\eta}. \label{kappaeq}
\end{align}
\\
Here,
\begin{align}
\begin{split}
{}^{\bar{\kappa}}\!A_{pqrs} &= 2\delta_{pr}\delta_{qs}(\epsilon_p-\epsilon_q) \\
&\quad+ (v_s - v_r)(8
g_{pqrs}-2
g_{prqs}-2
g_{psrq})
\end{split}
\end{align}
is the Hartree-Fock Hessian, with $v_r$ denoting the occupancy (0 or 1) of orbital $r$ in the Hartree-Fock state.\cite{MP:derivatives} 
In the case of CCSD, $\bm{^{\bar{\kappa}}\!\eta}$ is given as
\cite{MP:derivatives,Hald_02}
\begin{align}
\begin{split}
    ^{\bar{\kappa}}\eta_{pq} &= \sum_t h_{pt} (D_{tq} + D_{qt}) - \sum_t h_{qt} (D_{tp} + D_{pt})\\
    &\quad+\sum_{rst} g_{ptrs} (d_{tqrs} + d_{qtrs}) - \sum_{rst} g_{qtrs} (d_{tprs} + d_{ptrs}).
\end{split}
\end{align}
Similarly, the amplitude multipliers $\bar{\zeta}_\mu$ are determined from the stationarity condition {with respect to the coupled cluster amplitudes} $t_\mu$,
\begin{align}
    \dfrac{\partial \mathcal{L}}{\partial t_{\mu}}&=0\label{stattbar},
\end{align}
which yields different equations for the ground and the excited states. In particular,
\begin{align}
      \bm{\bar{\zeta}}^{\mathrm{GS}}\bm{A}&=-\bm{\eta}\label{eq:tbargs}\\
    \bm{\bar{\zeta}}^{\mathrm{ES}}\bm{A}&=-\bm{\eta}-\bm{L}\bm{A}R_0-\bm{J}\bm{A}-\bm{F(L)}\bm{R},
    \label{eq:tbares}  
\end{align}
with
\begin{align}
A_{\mu\nu} &= \bra{\mu}[\bar{H},\tau_{\nu}]\ket{\hf} \\
    \eta_{\mu} &= \bra{\hf}[\bar{H},\tau_{\mu}]\ket{\hf}\\
    J_{ai} &= \sum_{bj} L_{ij}^{ab}R_j^b\\
    F(L)_{\mu\nu} &= \sum_{\lambda}L_{\lambda}\bra{\lambda}[[\bar{H},\tau_{\mu}],\tau_{\nu}]\ket{\hf}.
\end{align}
For the ground state, we obtain the ground state multiplier equations, Eq.~\eqref{eq:tbargs}, and we will let $\bm{\bar{\zeta}}^{\mathrm{GS}} = \bm{\tbar}$, following the conventional notation for these multipliers.\cite{helgaker2014molecular}
For the excited states, Eq.~\eqref{eq:tbares} is obtained. The excited state multipliers, referred to below as the amplitude response, will similarly be denoted as $\bm{\bar{\zeta}}^{\mathrm{ES}} = \bm{\tbar}^{\textrm{ES}}$. 

With all orbital and wave function parameters variationally determined, 
the gradient can now be 
evaluated
as
the first {partial} derivative of the Lagrangian 
with respect to the nuclear 
coordinates.
As is well known, this derivative
 can be written as\cite{Helgaker1992,Hald_02}
\begin{align}
    \mathcal{L}^{(1)} = \sum_{pq} h_{pq}^{(1)}D_{pq} + \sum_{pqrs} g_{pqrs}^{(1)}d_{pqrs} + \sum_{p\leq q} \bar{\kappa}_{pq}F_{pq}^{(1)},
    \label{eq:grad}
\end{align}
where $h_{pq}^{(1)}$ and $g_{pqrs}^{(1)}$ denote one- and two-electron derivative integrals
and 
where
$F_{pq}^{(1)}$ 
denotes
the derivative Fock matrix. 
The one- and two-electron derivative integrals are here evaluated in an orthonormal MO (OMO) basis, {i.e.},
a basis strictly orthonormal at all geometries.~\cite{HELGAKER1988183,Helgaker2012} 
These OMOs are
obtained from the nonorthogonal unmodified MOs (UMOs), which are defined 
from 
the AOs at the displaced geometry and the MO coefficients of the unperturbed geometry.
The 
orthonormalization 
matrix that transforms 
UMOs to 
OMOs defines an orbital connection and is known as the connection matrix.\cite{Helgaker2012}

Here we will use
the symmetric connection, 
for which
the connection matrix is 
given as
the inverse square root of the UMO overlap matrix.\cite{HELGAKER1988183}
For this connection,
the derivative of the OMO Hamiltonian 
can be written as
\begin{align}
    H^{(1)} = H^{[1]} - 
    \frac{1}{2}\{S^{[1]}, H \}, 
    \label{eq:reortho_symcon}
\end{align}
where $H^{[1]}$ and $S^{[1]}$ denote derivatives of the Hamiltonian and of the overlap in the UMO basis. 
The second term of Eq.~\eqref{eq:reortho_symcon} is known as the reorthonormalization term. 
The
UMO
derivatives are evaluated by differentiating the AO integrals and then transforming them to the UMO basis.
The notation $\{ A, B \}$ 
means
the sum of all one-index transformations of $A$ and $B$. 
The expression for $F^{(1)}$ is similar to that for $H^{(1)}$ in Eq.~\eqref{eq:reortho_symcon} and is omitted.
See Refs.~\citenum{HELGAKER1988183} and \citenum{Helgaker2012} for further details on orbital connections. 

We will 
begin by deriving
expressions for the UMO contributions to the gradient, that is, the contributions originating from the first term in Eq.~\eqref{eq:reortho_symcon}. The reorthonormalization contributions are presented separately, see Section \ref{sec:reortho}.

The first term in Eq.~\eqref{eq:grad} has been thoroughly described 
in other works for the CCSD case, e.g.~by Scheiner \emph{et al.}\cite{Scheiner_87} The second and third terms
are not trivial and will be described in more detail.
For the second term, the CCSD two-electron densities are required. 
Expressions for both ground and excited state densities have been rederived 
and are given in Appendix \ref{sec:densities}.
Evaluating this term also requires that we consider the electron repulsion integrals.
Below, unless otherwise specified, these integrals are expressed in the $T_1$-transformed basis, and hence the densities have been made independent of $T_1$; see Eq.~\eqref{eq:density_general}. 
In the $T_1$-transformed basis, the Hamiltonian integrals can be written 
\begin{align}
    h_{pq} &= \sum_{rs} x_{pr} y_{qs} h_{rs}^\mathrm{MO} \\
    g_{pqrs} &= \sum_{tumn} x_{pt} y_{qu} x_{rm} y_{sn} g_{tumn}^\mathrm{MO},
\end{align}
where $\boldsymbol{x} = \boldsymbol{I} - \boldsymbol{t}_1$ and $\boldsymbol{y} = \boldsymbol{I} + \boldsymbol{t}_1^T$ and where $\boldsymbol{h}^\mathrm{MO}$ and $\boldsymbol{g}^\mathrm{MO}$ denotes integrals expressed in the MO basis.\cite{helgaker2014molecular}

To find expressions for the derivatives of $g_{pqrs}$, we expand the integral matrix in terms of its Cholesky decomposition. That is, we write
\begin{align}
\begin{split}
    g_{pqrs} = ( pq \vert rs ) &= \sum_{KL}(pq \vert K ) (\b{\mathcal{S}}^{-1})_{KL} (L \vert rs) \\
    &= \sum_{J}L_{pq}^J L_{rs}^J, 
\end{split}
\end{align}
where
\begin{align}
    \mathcal{S}_{KL} = ( K \vert L )
\end{align}
and where $K$ and $L$ are elements in the Cholesky basis.\cite{beebe_77} The Cholesky decomposition of $\b{\mathcal{S}}$ defines the $\b{Q}$ matrix, from which we can evaluate the inverse of $\b{\mathcal{S}}$:
\begin{align}
    \b{\mathcal{S}} = \b{Q} \b{Q}^T \implies \b{\mathcal{S}}^{-1} = 
    \b{Q}^{-T}
    \b{Q}^{-1}, \quad \b{Q}^{-T} = (\b{Q}^{-1})^T
\end{align}
This yields the definition of the Cholesky vectors:
\begin{align}
    L_{pq}^J = \sum_{K}(pq \vert K) Q^{-T}_{KJ}. \label{eq:choleskyvec}
\end{align}
The Cholesky vectors are also expressed in the $T_1$-transformed basis, where
\begin{align}
    L_{pq}^J = \sum_{rs} x_{pr} y_{qs} (L_{rs}^J)^\mathrm{MO}.
\end{align}
Here $(L_{rs}^J)^\mathrm{MO}$ denotes the Cholesky vectors in the MO basis.
From the above definitions, we can write the derivative two-electron integrals as
\begin{align}
\begin{split}
      (p q | r s)^{[1]}  &= \sum_{K}(p q \vert K )^{[1]} Z_{r s}^K +  \sum_{L}(r s \vert  L)^{[1]} Z_{p q}^L \\
      &\quad- \sum_{MN}Z_{p q}^M \mathcal{S}^{[1]}_{MN} Z_{r s}^N, 
\end{split}
\end{align}
where we have defined 
\begin{align}
    Z_{r s}^K &= \sum_L (\b{\mathcal{S}}^{-1})_{KL} (L \vert rs)
\end{align}
and used the identity
\begin{align}
    (\b{\mathcal{S}}^{-1})^{[1]} &= - \b{\mathcal{S}}^{-1} \b{\mathcal{S}}^{[1]} \b{\mathcal{S}}^{-1}
\end{align}
Upon contraction with the two-electron density, the second term of Eq. \eqref{eq:grad} becomes
\begin{align}
\begin{split}
    \sum_{pqrs} d_{pqrs} (pq \vert rs)^{[1]} 
    &= 2\sum_{pqK} (p q \vert K )^{[1]} W_{pq}^K - \sum_{MN}V_{MN} \mathcal{S}^{[1]}_{MN}\label{eq:integral_contraction}
\end{split}
\end{align}
with
\begin{align}
W_{pq}^K &= \sum_{rs}d_{pqrs} Z_{r s}^K\\
V_{MN} &= \sum_{pq}Z_{p q}^M W_{p q}^N.
\end{align}
The first term in Eq.~\eqref{eq:integral_contraction} is more conveniently calculated in the non-transformed basis by transferring the $T_1$-terms back to $W_{pq}^K$,
\begin{align}
    (W_{pq}^K)^{\mathrm{MO}}=\sum_{rs}x_{rp}y_{sq}W_{rs}^K,
\end{align}
before contracting with the differentiated MO integrals, $(pq|K)^{[1]\mathrm{,MO}}$.

\subsection{Two-electron density intermediates} \label{sec:intermediates}
Expressions for the various blocks of the two-electron density are reported in Appendix \ref{sec:densities}. 
In this section we describe in detail the two- and three-index density intermediates. All contributions to $W^J_{pq}$ from the $O^4$, $O^3V$, and $O^2V^2$ density blocks are constructed straight-forwardly by contracting the density block with $Z^J_{pq}$; hence, we will not discuss them further.
To avoid storing the $V^4$ and $OV^3$ 
blocks of the density
in memory, and, in addition, to avoid batching when constructing the $OV^3$ terms, 
we directly build their contributions to $W^J_{pq}$ and store these instead.
For improved readability, Einstein's implicit summation over repeated indices will be used in the remainder of this section. 
The contributions from the $OV^3$-density blocks  
to the gradient are 
\begin{align}
\begin{split}
    d_{abci} (ab \vert ci)^{[1]} &= 
    (a b \vert K )^{[1]} W_{ab}^K \\
    &\quad + (c i \vert  L)^{[1]} W_{c i}^L - V_{M N} \mathcal{S}^{[1]}_{MN} 
\end{split}
    \\
\begin{split}
     d_{abic} (ab \vert ci)^{[1]} &=
    (a b \vert K )^{[1]} W_{ab}^K \\
    &\quad+ (i c \vert  L)^{[1]} W_{i c}^L - V_{M N} \mathcal{S}^{[1]}_{MN}.
\end{split}
\end{align}
We thus 
directly construct the contributions to $W_{ci}^K$ and $W_{ic}^K$,
as well as to $W_{ab}^K$ and $V_{MN}$, 
from the two $OV^3$ blocks of the two-electron density.
From $d_{abci}$ we get the contributions
\begin{align}
    W_{ab}^K &= R_j^b X_{aj}^K\\
    W_{ci}^K &= L_{ji}^{ac}P_{aj}^K\\
    V_{MN} &= Z_{a b}^M W_{a b}^N + Z_{ci}^M W_{ci}^N,
\end{align}
where $L_{ji}^{ac}$ and $R_j^b$ denote excited state amplitudes, see Appendix \ref{sec:densities}. From $d_{abic}$ we similarly obtain
\begin{align}
\begin{split}
    W_{ab}^K
    &= 
    L^{a}_{m}\Tilde{Y}_{bm}^{K}(U)
    \\
    &\quad+
    2X_{ab}(L_2,T_2)P^K-O^J_{ai}R_i^{b}\\
    &\quad-(V_{akbi}(L_2,T_2)+Y_{akbi}(L_2,T_2))P^K_{ik}+2Q_{ak}^KR_k^{b}\\
    &\quad+W_{am}\Tilde{Y}_{bm}^K(T_2)
    -K_{icab}Z_{ic}^K
\end{split}
    \\
\begin{split}
    W_{ic}^K 
    &=
    \Tilde{t}^{cb}_{im}R_0 O_{bm}^K+
    (2\Tilde{R}_{mi}^{bc}
    -\Tilde{R}_{mi}^{cb})O_{bm}^K\\
    &\quad+
    2C^KR_{i}^{c}-X_{ac}(L_2,T_2)P_{ai}^K\\
    &\quad-K_{ik}^K R_k^{c}+2Y_{akci}(L_2,T_2)P_{ak}^K\\
    &\quad+U_{bm}^K\Tilde{t}_{mi}^{bc}
    -S_{iacb}Z_{ab}^K
\end{split}
    \\
    V_{MN} &= Z_{a b}^M W_{a b}^N + Z_{ic}^M W_{ic}^N
\end{align}
The intermediates introduced in these contributions are given in Table \ref{tab:intermediates}.
\begin{table}[htb]
\caption{Intermediates used in constructing the $W_{ab}^K$, $W_{ci}^K$, and $W_{ic}^K$ density intermediates. Observe that $C_2$ denotes a set of double amplitudes, and $\tilde{C}_{ij}^{ab} = 2 C_{ij}^{ab} - C_{ji}^{ab}$. In the expressions given in this table, $C_2 = T_2$. The definition of $C_2$ may be different as $C_2$ is merely a placeholder; see also Appendix \ref{sec:densities}. 
Summation over repeated indices is assumed. 
}
    \begin{ruledtabular}
    \centering
    \begin{tabular}{l} \\
    $\tilde{t}^{ab}_{ij}=2t^{ab}_{ij} - t^{ab}_{ji}$ \vspace*{0.3cm}\\
    $W_{ai} = L_{il}^{ad} R_{dl}$\vspace*{0.3cm}\\
    $X_{ab}(L_2,C_2) = L_{kl}^{ad} C_{kl}^{bd}$ \vspace*{0.3cm}\\
    $Y_{ajbi}(L_2,C_2) = L_{jl}^{ad} C_{il}^{bd}$ \vspace*{0.3cm}\\
    $V_{ajbi}(L_2,C_2) = L_{kj}^{ac} C_{ki}^{bc}$\vspace*{0.3cm}\\
    $U = T_2 R_0 + R_2$\vspace*{0.3cm}\\
    $S_{iacb} = B_{iamn}t_{mn}^{bc}$\vspace*{0.3cm}\\
    $B_{iamn} =L_{mn}^{ad}R_{i}^{d}$\vspace*{0.3cm}\\
    $K_{ik}^K = (V_{akbi}(L_2,T_2)+Y_{akbi}(L_2,T_2))Z_{ab}^K$\vspace*{0.3cm}\\
    $X_{aj}^K = L_{ji}^{ac}Z_{ci}^K$  \vspace*{0.3cm}\\
    $P_{aj}^K = R_j^bZ_{ab}^K$ \vspace*{0.3cm}\\
    $\Tilde{Y}_{bm}^{K}(C_2) = \Tilde{C}^{cb}_{im}Z_{ic}^K$\vspace*{0.3cm}\\
    $P^K =R_{i}^{c}Z_{ic}^K$\vspace*{0.3cm}\\
    $O^J_{ai} = X_{ac}(L_2,T_2)Z_{ic}^K$\vspace*{0.3cm}\\
    $P^K_{ik} = R_k^{c}Z_{ic}^K$\vspace*{0.3cm}\\
    $Q_{ak}^K = Y_{akci}(L_2,T_2)Z_{ic}^K$\vspace*{0.3cm}\\
    $K_{icab} = (L_{mn}^{ad}R_{i}^{d})t_{mn}^{bc}$\vspace*{0.3cm}\\
    $Q^K = X_{ab}(L_2,T_2)Z_{ab}^K$\vspace*{0.3cm}\\
    $P^K_{ai} = R_i^{b}Z_{ab}^K$\vspace*{0.3cm}\\
    $U_{bm}^K = W_{am}Z_{ab}^K$ \vspace*{0.3cm}
    \end{tabular}
    \end{ruledtabular}
    \label{tab:intermediates}
\end{table}

The $V^4$ density contribution to $W^J_{ab}$ is also evaluated directly as a contraction between the density and $Z_{cd}^K$, albeit with batching. This contribution is the steepest-scaling term in the gradient and 
implies the calculation of the density contribution
\begin{align}
    O^{(2)}_{abcd} &= L_{ij}^{ac} C_{ij}^{bd}. \label{eq:v4density}
\end{align}
To evaluate this term 
as written would have 
a cost of $O^2V^4$. However, it is possible to reduce the cost by a factor of four by adapting the well-known strategy for constructing the A2 term of $\bm{\Omega}$.\cite{Scuseria1988} In the case of Eq.~\eqref{eq:v4density}, we form the symmetric and anti-symmetric combinations of $L_2$ and $C_2$:
\begin{align}
    X_{ij}^{ab\pm} &= X_{ij}^{ab} \pm X_{ji}^{ab}, \quad X = L_2, C_2.
\end{align}
Then, by contracting the symmetric and anti-symmetric terms separately and adding them together, we need not loop over all indices, 
but merely $i\geq j$, $a\geq c$, and $b\geq d$, leading to an eight-fold reduction in cost. However, since this must be done twice (for symmetric and anti-symmetric terms), the net reduction in cost is a factor of four.

\subsection{Orbital relaxation contributions}

In order to obtain the gradient, all three terms of Eq.~\eqref{eq:grad} must be evaluated. So far, we have not yet discussed the third term.
First, the $\bar{\kappa}$ parameters are determined from the 
Z-vector equation given in Eq.~\eqref{kappaeq}.
Since CCSD is orbital invariant, 
we only consider the
VO block of the $\bar{\kappa}$ vector.\cite{Hald_02}  
The UMO contribution to the orbital relaxation then reads
\begin{align}
\begin{split}
    \sum_{ai}\bar{\kappa}_{ai}&F_{ai}^{[1]} = \sum_{ai}\bar{\kappa}_{ai}\Big(h_{ai}^{[1]} +\sum_{j}(2g_{aijj}^{[1]}-g_{ajji}^{[1]})\Big).\\
\end{split}
\end{align}
The integrals are here expressed in the MO basis. 
For efficiency, the second and third terms are rewritten by using the Cholesky decomposition:
\begin{align}
\begin{split}
    \sum_{aij}\bar{\kappa}_{ai}(2g_{aijj}^{[1]}-g_{ajji}^{[1]}) &=
      \sum_{aiK}K_{ai}^K(ai\mid K)^{[1]} \\
      &+ \sum_{jK}L^K (K\mid jj)^{[1]} +\sum_{ijK}N_{ji}^K(K\mid ji)^{[1]}  \\
    &- \sum_{KL}M_{KL} S_{KL}^{[1]} + \frac{1}{2}\sum_{KL}O_{KL} S_{KL}^{[1]}.
\end{split}
\end{align}
The introduced intermediates are given in Table \ref{tab:intermediates2}.
\begin{table}[htb]
\caption{Intermediates used in constructing the $\sum_{aij}\bar{\kappa}_{ai}(2g_{aijj}^{[1]}-g_{ajji}^{[1]})$. Here $Z_{pq}^K$ are expressed in the MO basis.
}
    \begin{ruledtabular}
    \centering
    \begin{tabular}{l}
    \\
       $K_{ai}^K = \sum_{j}\bar{\kappa}_{ai}(2Z_{jj}^K-Z_{ji}^K)$ \vspace*{0.3cm}\\ 
    $L^K = 2\sum_{ai}\bar{\kappa}_{ai}Z_{ai}^K$\vspace*{0.3cm}\\
    $N_{ji}^K = -\sum_{a}\bar{\kappa}_{ai}Z_{aj}^K$\vspace*{0.3cm}\\ 
    $M_{KL} = \sum_{j}L^KZ_{jj}^L$\vspace*{0.3cm}\\ 
    $O_{KL} = \sum_{ij} N_{ji}^KZ_{ji}^L$ 
    \vspace*{0.3cm}
    \end{tabular}
    \end{ruledtabular}
    \label{tab:intermediates2}
\end{table}
The first three terms are added to the $W_{pq}^Jg_{pqrs}^{[1]}$ term, while the remaining terms are added to the $V_{MN}S_{MN}^{[1]}$ term; see Eq.~\eqref{eq:integral_contraction}.

To determine $\bar{\kappa}_{ai}$, we also need the right-hand-side of the Z-vector equation.
This vector is conveniently constructed from a three-index intermediate $\tilde{W}_{pq}^J = W_{pq}^J(L)$, defined as $W_{pq}^J$, but constructed using $L_{pq}^J$ rather than $Z_{pq}^J$. In terms of this intermediate, and using integrals expressed in the MO basis, we have\cite{Hald_02}
\begin{align}
\begin{split}
    {^{\bar{\kappa}}}\eta_{ai} = (1 - P_{ai}) \Bigl( &\sum_t \mathcal{D}_{ti} 
    h_{at}
    + \sum_{tJ} \mathcal{\tilde{W}}_{ti}^J
    L_{at}^J \Bigr),
\end{split}
\end{align}
where $P_{ai} X_{ai} = X_{ia}$ and
\begin{align}
    \mathcal{D}_{pq} &= D_{pq} + D_{qp} \\
    \tilde{\mathcal{W}}_{pq}^J &= \tilde{W}_{pq}^J + \tilde{W}_{qp}^J.
\end{align}

\subsection{Reorthonormalization contributions} \label{sec:reortho}
The second term in Eq.~\eqref{eq:reortho_symcon} gives rise to reorthonormalization contributions to the gradient. 
These contributions
can 
be expressed as $- \sum_{pq} \mathcal{F}_{pq} S^{[1]}_{pq}$, where
\begin{align}
    \mathcal{F}_{pq} = \sum_i (\mathcal{D}_{pi} h_{qi} + \sum_J\tilde{\mathcal{W}}_{pi}^J L_{qi}^J) + \mathcal{F}_{pq}^{\bar{\kappa}}.\label{eq:reorthonorm_term}
\end{align}
The orbital relaxation contribution $\mathcal{F}_{pq}^{\bar{\kappa}}$ is described in detail in Appendix \ref{appendix_relaxation}. Note that all terms in Eq.~\eqref{eq:reorthonorm_term} are given in the MO basis.

\section{Computational details}
\begin{figure}
    \centering
    \includegraphics[width=\linewidth]{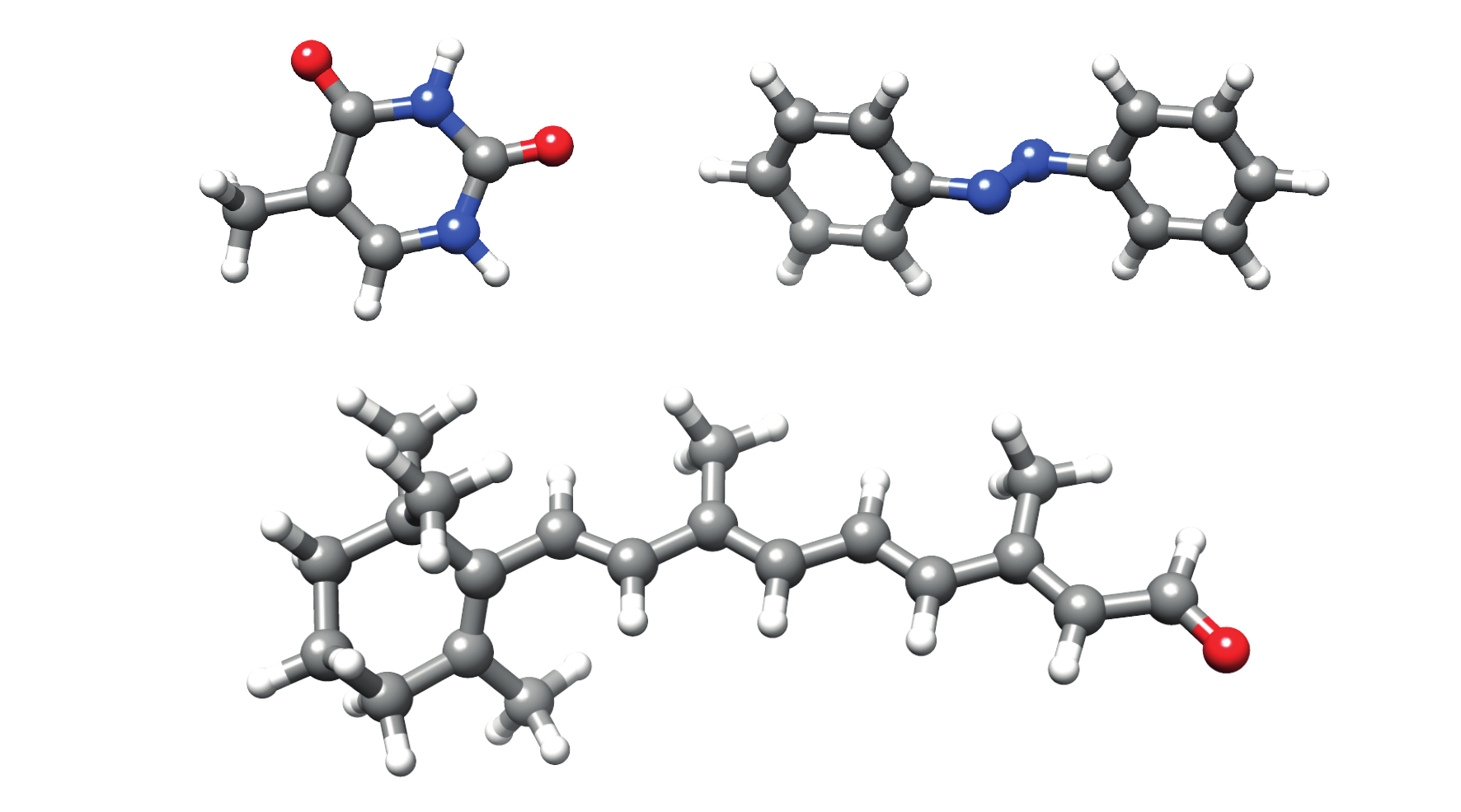}
    \caption{Ground state geometries, optimized at the CCSD/aug-cc-pVDZ level, for thymine (top left), azobenzene (top right), and retinal (bottom).}
    \label{fig:geoms}
\end{figure}

The CD-CCSD gradient for ground and excited states has been implemented in a development version of $e^T$ 1.4. We apply the gradient implementation
to determine equilibrium geometries 
 for thymine, azobenzene, and retinal (see Figure \ref{fig:geoms}), where we consider the ground state in all three systems 
 and
 the lowest singlet excited state in thymine and azobenzene. 
 To perform these calculations, we
 have also
  implemented an optimizer that uses the Broyden-Fletcher-Goldfarb-Shanno (BFGS) algorithm with the rational function (RF)\cite{banerjee1985search} level shift.
 This implementation makes use of the redundant internal coordinates introduced by Bakken and Helgaker,\cite{Bakken2002}
 along with the initial ``simple model Hessian guess'' proposed by the same authors.

For comparison, we carried out CD-CCSD  calculations with Q-Chem 5.4~\cite{Qchem541}
and
RI-CCSD calculations with Psi4 1.3.\cite{psi4}
An aug-cc-pVDZ basis was used in all calculations.
For consistency with our implementation, 
we disabled 
the frozen core approximation 
in Q-Chem and Psi4.
Default thresholds were used in all calculations, except in the case of the CD convergence threshold. The reason for this is that a CD threshold of e.g.~$10^{-3}$ can result in slow convergence because the Cholesky basis varies on the potential energy surface, causing discontinuities of the same order of magnitude as the CD threshold. This was also observed by
Feng \emph{et al}.\cite{Krylov_19}
Thus, in order to converge the gradient to $3 \times 10^{-4}$ (the Baker convergence criterion), we employed a tighter CD threshold of $10^{-4}$ throughout. The one exception to this was for
the large retinal molecule, 
where
we instead applied a CD threshold of $10^{-3}$. 

The $e^T$ program does not utilize point group symmetry, whereas this was enabled in Q-Chem and Psi4. However, the initial geometries does not possess point group symmetry, and this should therefore not
affect the comparison of timings significantly. 
Initial and optimized molecular geometries can be found in Ref.~\citenum{zenodo}. At the initial geometries, the excitation energies in thymine and azobenzene are 5.20 eV and 3.46 eV, respectively; at the 
optimized 
excited state
geometries, the excitation energies are 3.98 eV and 2.35 eV.
All calculations were performed on one  
node with two {Intel Xeon E5-2699 v4} processors with 44 cores and given 1 TB of shared memory. 

\section{Results and discussion}
\subsubsection{Timing Comparisons of Different Implementations}
To demonstrate the efficiency of our implementation, we compare calculation times for geometry optimizations of a small system (thymine) and a medium-large system (azobenzene) with the Psi4 and Q-Chem programs.
Excited state gradients are not available in Psi4 and thus excited state geometry optimizations have not been performed with this program. We report the percentage of the 
time spent 
determining the gradients for the $e^T$ calculations. This is not reported for Q-Chem or Psi4, as this time is not directly available from their respective outputs.

\begin{table*}[htb]
\caption{Ground state calculation times for thymine and azobenzene using Psi4 and our new implementation in $e^T$. The number of optimization cycles required, $n_\mathrm{cycles}$, are reported, as well as the total calculation time, $t_{\mathrm{total}}$, and the time per cycle, $t_\mathrm{cycle}$. The average fraction of time spent calculating the gradient per optimization cycle, $t_\mathrm{gradient}$, is reported for $e^T$.
}
    \begin{ruledtabular}
    \centering
    \begin{tabular}{lccccccc}
         &\multicolumn{4}{c}{$e^T$}&\multicolumn{3}{c}{Psi4}\\
         \cline{2-5}
         \cline{6-8}
         & $n_\mathrm{cycles}$ & $t_\mathrm{total}$ & $t_\mathrm{cycle}$ & $t_\mathrm{gradient}$ & $n_\mathrm{cycles}$ & $t_\mathrm{total}$ & $t_\mathrm{cycle}$ \\
         \hline 
         Thymine    & 6  & 45m     & 7m  & 7 \%  & 7              & 2h 6m      & 18m \\
         Azobenzene & 11 & 10h 26m & 57m & 8.9\% & 19$^{\dagger}$ & 42h 13m & 2h 13m \\
    \end{tabular}
    \end{ruledtabular}
{$^{\dagger}$ The calculation did not converge in 19 cycles in Psi4}
    \label{tab:timings_psi4}
\end{table*}

Timings for $e^T$ and Psi4 are given in Table \ref{tab:timings_psi4}.
From the results in the table 
we see
that the 
calculation
of the gradient 
amounts to
only 7\%--9\% of the total calculation time, illustrating the efficiency of 
the gradient
implementation. 
Here, the time required to determine the multipliers is not included in the gradient time.
We observe that the calculation time required by $e^T$, per optimization cycle, is roughly half of that required by Psi4.

When applying an inner projection method,
the integral costs 
are 
proportional to the size of the auxiliary basis.\cite{folkestad_19}
In the density-fitting scheme used in Psi4, the thymine calculation required 786 auxiliary basis functions, while 967 functions
were used
in our CD-based 
implementation. 
These two numbers are of the same order of magnitude;
thus, the 
computational resources
required
should be comparable. 
This is also the case for the azobenzene calculations, 
in which Psi4 and $e^T$ required 1238 and 1444 auxiliary basis functions, respectively.
Hence, the two approaches are similar in terms of computational costs. However, note that the time spent on the Cholesky decomposition itself is negligible and that the resulting integral errors are strictly lower than
the CD threshold (here $10^{-4}$). Such strict error control is not possible with the density-fitting method.\cite{folkestad_19}

\begin{table*}[hbt!]
\caption{Ground and excited state calculation times for thymine and azobenzene using Q-Chem and our new implementation in $e^T$. The excited states are the lowest singlet states in all cases. The number of optimization cycles required, $n_\mathrm{cycles}$, are reported, as well as the total calculation time, $t_\mathrm{total}$, and the time per cycle, $t_\mathrm{cycle}$. The average fraction of time spent calculating the gradient per optimization cycle, $t_\mathrm{gradient}$ is reported for $e^T$.\\
     }
    \label{tab:timings_qchem}
\begin{ruledtabular}
    \centering
    \begin{tabular}{lccccccc}
         &\multicolumn{4}{c}{$e^T$}&\multicolumn{3}{c}{Q-Chem}\\
         \cline{2-5}
         \cline{6-8}
         & $n_\mathrm{cycles}$ & $t_\mathrm{total}$ & $t_\mathrm{cycle}$ & $t_\mathrm{gradient}$ & $n_\mathrm{cycles}$ & $t_\mathrm{total}$ & $t_\mathrm{cycle}$ \\\hline
         Thymine (GS)& 6 & 45m & 7m & 7 \% & 10 & 6h 22m & 38m \\
         Thymine (ES)& 7 & 2h 51m & 25m & 4.5\% & 7 & 7h 01m   & 1h \\
        Azobenzene (GS) & 11 & 10h 26m & 57m & 8.9\% & 7 & 26h 08m & 3h 44m
        \\
         Azobenzene (ES) & 10 & 47h 43m & 4h 46m & 4.4\% & 6 & 37h 40m & 6h 17m \\
    \end{tabular}
\end{ruledtabular}
\end{table*}

Timings for $e^T$ and Q-Chem are given in 
Table~\ref{tab:timings_qchem}.
We now also consider excited state geometry optimizations, and observe once again that the calculation time is not dominated 
by the gradient.
In fact, 
the calculation of the gradient amounts to 5\%  of the full calculation time for the excited state (compared to 10\% for the ground state). Note
that when reporting the time spent on determining the excited state gradients, we have again excluded the time needed to determine the amplitude response. 

From Table~\ref{tab:timings_qchem} it is furthermore observed that 
we can carry out
ground state optimizations
in roughly 50\% of the time required by \mbox{Q-Chem}, or even less. 
The savings are significantly larger for 
ground state optimizations. 
This
can be attributed to differences, between the two programs, in the efficiency of the ground and excited state implementations. 
The calculation time in \mbox{Q-Chem} can be roughly halved with the utilization of the frozen core approximation (for timings, see SI), resulting in computation times for the excited state that are roughly the same as those observed in $e^T$ without this approximation. For the ground state, however, our new implementation still offers significant time savings, 
despite frozen core calculations being inherently less computationally demanding.

The reported comparisons were carried out without enforcing point-group symmetry. As also noted above, further improvements to the Q-Chem and Psi4 timings could have been obtained by starting from geometries with point-group 
symmetry.

\subsubsection{Convergence Threshold Effects}
In addition to investigating the calculation times, we also studied the effect of changing the CD and gradient convergence thresholds on the final geometry of thymine---as measured by changes in redundant internal coordinates.\cite{Bakken2002} From Table~\ref{tab:thresh} we observe that a very small error is obtained in the final geometry, even at a CD threshold of $10^{-3}$. This 
small error
was also pointed out by Feng \emph{et al.}\cite{Krylov_19} 
At our chosen CD threshold of $10^{-4}$, the largest change in bond length is less than $0.01$ pm; the changes in angles and dihedral angles are 
smaller than 0.002 radians, corresponding to about 0.01$^{\circ}$. Thus, at the CCSD level of theory, the optimized geometry of thymine can be considered fully converged with our chosen 
CD threshold.

\begin{table*}[htb!]
 \caption{Deviations in the optimized geometry of thymine. The deviations are given relative to an optimized geometry obtained with a CD convergence threshold of $10^{-8}$ and a gradient convergence threshold of $3\cdot 10^{-8}$. 
 Changes in internal coordinates are measured by the largest change in a bond length ($\Delta r$), an angle ($\Delta \alpha$), and a dihedral angle ($\Delta \theta$).}
    \label{tab:thresh}
\begin{ruledtabular}
    \centering
    \begin{tabular}{ccccc}
         CD threshold & Gradient threshold &$\textrm{Max }\Delta r$/pm&$\textrm{Max }\Delta \alpha$/rad&$\textrm{Max }\Delta \theta$/rad\\\hline
           $10^{-3}$&$3\cdot 10^{-3}$&0.043&0.00062&0.0012\\
            $10^{-4}$&$3\cdot 10^{-4}$&0.006&0.00032&0.0012\\
            $10^{-5}$&$3\cdot 10^{-5}$&0.001&0.00002&0.0010\\
    \end{tabular}
\end{ruledtabular}
\end{table*}
It must be noted, however, that even at this threshold, small differences in the CD basis can occur. This may cause changes in the number of required optimization cycles from run to run, but
the final geometries remain unchanged within the specified convergence thresholds.
\subsubsection{Illustration of Large-Scale Application}
To showcase the applicability of our implementation, we have also performed a calculation on retinal, which contains 49 atoms and 150 electrons, corresponding to 735 basis functions with our chosen basis set (aug-cc-pVDZ). This calculation was only carried out using $e^T$. 
\begin{table}[htb]
\caption{Calculation time for optimizing the geometry of retainal in $e^T$. The number of optimization cycles required, $n_\mathrm{cycles}$, are reported, as well as the total calculation time, $t_\mathrm{total}$, the time spent per cycle, $t_\mathrm{cycles}$, and the average time spent calculating the gradient per optimization cycle, $t_\mathrm{gradient}$. The CD convergence threshold was here $10^{-3}$.}
\label{tab:timings_retinal}
\begin{ruledtabular}
    \centering
    \begin{tabular}{ccccc}
         &$n_\mathrm{cycles}$&$t_\mathrm{total}$ & $t_\mathrm{cycle}$ & $t_\mathrm{gradient}$\\\hline
         Retinal (GS) & 24 & 17d 10h 23m & 17h 26m & 6.6\% \\
    \end{tabular}
\end{ruledtabular}
\end{table}
While 
the calculation does indeed put a strain on the computational resources, requiring 17 hours per optimization cycle, it can in fact be performed. 
As in the other calculations, 
the time spent calculating the gradient amounts to only a small fraction of the total calculation time.

\section{Conclusions}
We have presented an efficient implementation of CCSD gradients for 
ground and excited states based on Cholesky-decomposed electron repulsion integrals. 
Since CD is an inner projection scheme, we have chosen an implementation approach that follows earlier schemes\cite{Bozkaya_16,Delcey2014} for CD and RI
where one avoids the storage of $OV^3$ and $V^4$ arrays by constructing 3-index 
intermediates. 
We have furthermore chosen not to store the derivative Cholesky vectors; instead,  the associated contributions are constructed on-the-fly. This allows us to significantly reduce 
the 
storage requirements.

Relative to Psi4 and Q-Chem, our implementation was shown to reduce the calculation time of geometry optimizations by roughly a factor of two.
To a large extent, this reduction in time reflects the efficiency of the coupled cluster code in the $e^T$ program.
However,
the calculation of gradients 
was found to require 
only a small fraction of the total calculation time, showcasing 
the efficiency of the gradient implementation. 

The capabilities of our implementation was highlighted by showing that a geometry optimization of retinal could be carried out.

Further reduction in computational demands would be achieved by means of the frozen-core approximation. Work on this is currently in progress.

\section{Supplementary material}
Comparison of frozen-core and non-frozen-core calculations in Q-Chem.

\section{Data availability statement}
Geometries can be found in Ref.~\citenum{zenodo}. The code will be released in an upcoming version of the $e^T$ program, which is open-source.\cite{etpaper} 

\begin{acknowledgments}
We acknowledge support from the DTU Partnership PhD programme (PhD grant to AKSP).
AKSP acknowledges funding from the European Cooperation in Science and Technology, COST Action CA18222 {\it Attochem}.
SC acknowledges support from 
the Independent
Research Fund Denmark (DFF-RP2 Grant 7014-00258B).
E.F.K, S.C., and H.K. acknowledge the
Research Council of Norway through FRINATEK projects
263110 and 275506. 
Computing resources through
UNINETT Sigma2—the National Infrastructure for High
Performance Computing and Data Storage in Norway
(Project No. NN2962k) are also acknowledged.
\end{acknowledgments}

\appendix
\section{Two-electron densities}\label{sec:densities}
The two-electron density is here taken as
\begin{align}
    d_{pqrs} = L_{\nu}\Tbraket{\nu}{e^{-T_2} e_{pqrs} e^{T_2}}{\rho}R_{\rho}, \label{eq:density_general}
\end{align}
where 
 \begin{align}
    e_{pqrs} = E_{pq} E_{rs} - \delta_{qr} E_{ps}.
\end{align}
Note that we are using a $T_1$-transformed basis, where the $T_1$-dependence has been moved into the derivative integrals, as they will be contracted with the density later on. Throughout, we assume a spin-adapted singlet basis, where the kets are expressed in the so-called elementary basis and bras are expressed in the basis biorthonormal  to the kets.\cite{helgaker2014molecular}

Recall the special cases in equation of motion theory; the ground state is described by
\begin{align}
    L_0 &= 1, \quad L_\mu = \Bar{t}_\mu, \\
    R_0 &= 1, \quad R_\mu = 0,
\end{align}
and the excited states by
\begin{align}
    L_0 &= 0, \quad\quad\quad\; L_\mu = L_\mu, \\
    R_0 &= -\b{\Bar{t}}^T \b{R}, \quad R_\mu = R_\mu. 
\end{align}
Here $\b{R}$ and $\b{L}$ are defined for $\mu > 0$, and
\begin{align}
    \b{A} \b{R} &= \omega \b{R} \\
    \b{A}^T \b{L} &= \omega \b{L}.
\end{align}
In addition, for the excited state, terms with 
\begin{align}
        L_0 &= 0, \quad L_\mu = \bar{t}^{\textrm{ES}}_\mu \\
    R_0 &= 1, \quad R_\mu = 0
\end{align}
must be determined to account for the terms of the excited state Lagrangian containing the the amplitude response $\bar{\zeta}_{\mu}$. These terms will however be formally identical to the terms of the ground state that do not include $L_0$, and these will therefore not be written out explicitly. This applies to both the one- and two-electron densities.

In the following, we shall utilize that $R_2$ can be written as $\Tilde{R}_2$:
\begin{align}
\begin{split}
    R_2 &= \frac{1}{2} \sum_{aibj} R_{aibj} (1 + \delta_{ai,bj}) E_{ai}E_{bj} \\
    &\equiv \frac{1}{2} \sum_{aibj} \Tilde{R}_{aibj} E_{ai}E_{bj}=\Tilde{R}_2.
\end{split}
\end{align}
There are eight unique combinations of occupied and virtual indices, because of the symmetry
\begin{align}
    e_{pqrs} = e_{rspq} \implies d_{pqrs} = d_{rspq}.
\end{align}
For each of these combinations, we derive below the corresponding CCSD two-electron density block.
For improved readability, Einstein's implicit summation over repeated indices will be used.

For the ground state we obtain and implement
\begin{align}
d_{ijkl}^{\textrm{gs}} &= (L_0 R_0) \Lambda_{ijkl} + O^{(2)}_{ijkl}(L_2, T_2) R_0 \\
d_{aijk}^{\textrm{gs}} &= (2 L^a_i \delta_{jk} - L^a_k \delta_{ji}) R_0 \\
d_{ijka}^{\textrm{gs}} &= O^{(M)}_{ijka}(L_1, T_2)R_0\\
d_{abij}^{\textrm{gs}} &= O^{(2)}_{abij}(L_2,T_2)R_0 \\
d_{aibj}^{\textrm{gs}} &= L_{ij}^{ab} R_0\\
d_{iajb}^{\textrm{gs}} &=2R_0L_0\Tilde{t}_{ij}^{ab}+ R_0\frac{1}{2}O_{iajb}^{(2)}(L_2,T_2)
    \\
    d_{aijb}^{\textrm{gs}} &= O^{(M)}_{aijb}(L_2,T_2)R_0\\
    d_{abci}^{\textrm{gs}} &= 0\\
    d_{abic}^{\textrm{gs}} &= L^{a}_{m}\Tilde{t}^{cb}_{im}R_0 \\
    d_{abcd}^{\textrm{gs}} &= O^{(2)}_{abcd}(L_2, T_2)R_0.
\end{align}
and,
for the excited state, we similarly obtain and implement
\begin{align}
\begin{split}
d_{ijkl}^{\textrm{es}} &= O^{(2)}_{ijkl}(L_2, U) + (\b{L}^T\b{R})\Lambda_{ijkl} \\
&\quad+ O_{ijkl}^{[1]}(L_1, R_1) 
\end{split}
\\
\begin{split}
d_{aijk}^{\textrm{es}} &= (2 L^a_i \delta_{jk} - L^a_k \delta_{ji}) R_0 - L_{ik}^{ad} R_{dj} \\
&\quad+ 2  W_{ai} \delta_{jk} - W_{ak} \delta_{ji}
\end{split}
\\
\begin{split}
 d_{ijka}^{\textrm{es}} &= O^{(M)}_{ijka}(L_1, U)\\
    &\quad-2X_{ij}(L_2,T_2)R_k^a\\
    &\quad+X_{kj}(L_2,T_2)R_i^a+X_{da}R_i^d\delta_{jk}\\
    &\quad+\Big(V_{djai}(L_2,T_2)+Y_{djai}(L_2,T_2)\\
    &\quad-2\delta_{ij}X_{da}(L_2,T_2))\Big)R_k^d\\
    &\quad+\Big(Z_{ijkl}(L_2,T_2)+X_{il}(L_2,T_2)\delta_{jk}\\
    &\quad-2\delta_{ij}X_{kl}(L_2,T_2))\Big)R_l^a\\
    &\quad-\tilde{t}_{ik}^{ea}W_{ej}-\Big(\tilde{t}_{mi}^{ea}\delta_{jk}-2\delta_{ij}\tilde{t}_{mk}^{ea}\Big)W_{em}\\
    &\quad-\tilde{Y}_{djak}(L_2,T_2)R_i^d
\end{split}
    \\
    d_{abij}^{\textrm{es}} &= O^{(2)}_{abij}(L_2,U) + O^{(1)}_{abij}(L_1,R_1)\\
    d_{aibj}^{\textrm{es}}&= L_{ij}^{ab} R_0\\
\begin{split}
    d_{iajb}^{\textrm{es}} &= O_{iajb}^{(2)}(L_2,U_2)\\
    &\quad+2L_{ck}\Tilde{t}_{ik}^{ac}R_{bj}-L_{ck}\Tilde{t}_{jk}^{ac}R_{bi}+2L_{ck}\Tilde{t}_{jk}^{bc}R_{ai}\\
    &\quad-L_{ck}\Tilde{t}_{ik}^{bc}R_{aj}
    -Y_{cb}\Tilde{t}_{ij}^{ac}-Y_{ca}\Tilde{t}_{ji}^{bc}-Y_{jk}\Tilde{t}_{ik}^{ab} \\
    &\quad-Y_{ik}\Tilde{t}_{kj}^{ab}+2 L_{ck}\Tilde{t}_{ij}^{ab}R_{ck} + {L_{kl}^{cd}\Tilde{R}_{kl}^{cd}\Tilde{t}_{ij}^{ab}}   
\end{split}
\\
   d_{aijb}^{\textrm{es}} &= O^{(M)}_{aijb}(L_2,U)
    + 2L_{ai}R_{bj}-Y_{ab}\delta_{ij}\\
    d_{abci}^{\textrm{es}} &= L_{ji}^{ac} R_j^b\\
\begin{split}
    d_{abic}^{\textrm{es}} &= L^{a}_{m}\Tilde{t}^{cb}_{im}R_0 \\
    &\quad+L_m^a(2\tilde{R}_{mi}^{bc}-\tilde{R}_{mi}^{cb}) \\
    &\quad+2X_{ab}(L_2,T_2)R_{i}^{c}\\
    &\quad-(V_{akbi}(L_2,T_2)+Y_{akbi}(L_2,T_2))R_{k}^{c}\\
    &\quad+L_{mk}^{ad}\Tilde{t}_{mi}^{bc}R_{k}^{d}-L_{mn}^{ad}t_{mn}^{bc}R_{i}^{d}\\
    &\quad+{\Tilde{Y}_{akci}(L_2,T_2)}R_k^{b}-X_{ac}(L_2,T_2)R_i^{b}
    \end{split}
    \\
    d_{abcd}^{\textrm{es}} &= O^{(2)}_{abcd}(L_2, U).
\end{align}
Here we have defined the following quantities:
\begin{align}
\Lambda_{ijkl} &= 4 \delta_{ij} \delta_{kl} - 2 \delta_{il}\delta_{kj}\\
\begin{split}
    O_{ijkl}^{(1)}(L_1, R_1) &= -2 Y_{ij} \delta_{kl} + Y_{il} \delta_{kj} \\
    &\quad- 2 \delta_{ij} Y_{kl} + \delta_{il} Y_{kj} 
\end{split}
    \\
    Y_{kj} &= R_{ek} L_{ej} = R_k^e L_j^e\\
\begin{split}
    O_{ijkl}^{(2)}(L_2, C_2) &= - 2 \delta_{kl} X_{ij}(L_2, C_2) \\
    &\quad- 2 \delta_{ij} X_{kl}(L_2, C_2)\\
    &\quad+ \delta_{kj} X_{il}(L_2, C_2) + \delta_{il}
    X_{kj}(L_2, C_2) \\
    &\quad+ Z_{ijkl}(L_2, C_2) 
\end{split}
    \\
    X_{ij}(L_2, C_2) &= C_{mi}^{ef} L^{ef}_{mj} \\
    Z_{ijkl}(L_2, C_2) &= C_{ik}^{ef} L^{ef}_{jl}\\
    U_2 & = T_2 \frac{1}{2}R_0 + \tilde{R}_2\\
\begin{split}
     O^{(M)}_{ijka}(L_1, T_2)&=L_{em}\left(2\delta_{ij}\Tilde{t}_{km}^{ae}-\delta_{jk}\Tilde{t}_{mi}^{ea}\right)R_0\\
     &\quad-L_{ej}\Tilde{t}_{ik}^{ea}R_0
\end{split}
    \\
    \tilde{Y}_{djak}(L_2,T_2)&=L_{mj}^{ed}\tilde{t}_{mk}^{ea} \\
     Y_{ab} &= L_k^a R_k^b \\
    O^{(1)}_{abij}(L_1,R_1) &= 2 \delta_{ij} Y_{ab}(L_1,R_1) - L_j^a R_i^b \\
\begin{split}
    O^{(2)}_{abij}(L_2,C_2) &= 2 \delta_{ij} X_{ab}(L_2,C_2) - Y_{ajbi}(L_2,C_2) \\
    &\quad- V_{ajbi}(L_2,C_2)
\end{split}
    \\
\begin{split}
    O_{iajb}^{(2)}(L_2,C_2) &=
     \Tilde{Y}_{ckai}(L_2,T_2)\Tilde{C}_{kj}^{cb}\\
     &\quad+\Tilde{Y}_{ckbj}(L_2,T_2)\Tilde{C}_{ki}^{ca}
    \\
    &\quad+ Z_{ikjl}(L_2,T_2)C_{kl}^{ab}\\
    &\quad+Z_{imjn}(L_2,C_2)t_{mn}^{ab}\\
    &\quad-X_{ca}(L_2,T_2)\Tilde{C}_{ij}^{cb}\\
    &\quad-X_{ea}(L_2,C_2)\tilde{t}_{ij}^{eb}
    \\
    &\quad-X_{ik}(L_2,T_2)\Tilde{C}_{kj}^{ab}\\
    &\quad-X_{im}(L_2,C_2)\tilde{t}_{mj}^{ab}\\
    &\quad -X_{jk}(L_2,T_2)\Tilde{C}_{ik}^{ab}\\
    &\quad-X_{jm}(L_2,C_2)\tilde{t}_{im}^{ab}
    \\
    &\quad-X_{cb}(L_2,T_2)\Tilde{C}_{ji}^{ca}\\
    &\quad-X_{eb}(L_2,C_2)\tilde{t}_{ji}^{ea}\\
    &\quad-\Tilde{Y}_{embi}(L_2,C_2)t_{mj}^{ea}\\
    &\quad-Y_{emaj}(L_2,C_2)\Tilde{t}_{mi}^{eb}
    \\
    &\quad+Y_{ckbi}(L_2,C_2)t_{kj}^{ac} \\
    &\quad+Y_{ckbi}C_{kj}^{ac}\\
   &\quad+V_{ckaj}(L_2,T_2)C_{ik}^{cb}\\
   &\quad+V_{ckbi}(L_2,T_2)C_{kj}^{ac}
  \end{split}
   \\
 \begin{split}
   O^{(M)}_{aijb}(L_2,C_2)&=-L_{im}^{ae}C_{jm}^{eb}R_0\\
   &\quad+2Y_{aibj}(L_2,C_2)-\delta_{ij}X_{ab}(L_2,C_2)
  \end{split}
   \\
   O^{(2)}_{abcd}(L_2, C_2) &= L_{ij}^{ac} C_{ij}^{bd},
\end{align}
where $C_2$ denotes a set of double amplitudes and $\tilde{C}_2$ is defined in an identical fashion to $\tilde{t}$.

All $O^4$, $O^3V$ and $O^2V^2$ densities are directly constructed and stored in memory, while the $OV^3$ and $V^4$ densities are not explicitly constructed. As discussed in Section~\ref{sec:intermediates}, we instead construct their contributions to the 3-index density intermediates $W^J_{pq}$ and store these in memory.

\section{Orbital relaxation in reorthonormalization}
\label{appendix_relaxation}
The relaxation contributions---which we derive by expanding $h_{pq}^{(1)}$ and $g_{pqrs}^{(1)}$ using Eq.~\eqref{eq:reortho_symcon} and inserting into the differentiated Fock matrix, see Eq.~\eqref{eq:fockmatrix} and \eqref{eq:grad}---are
\begin{align}
    \mathcal{F}_{aq}^{\bar{\kappa}} &= \frac{1}{2} \bar{\kappa}_{ai} h_{qi} + (g_{qijj} - \frac{1}{2}g_{qjji}) \bar{\kappa}_{ai} \\
\begin{split}
    \mathcal{F}_{iq}^{\bar{\kappa}} &= \frac{1}{2} \bar{\kappa}_{ai} h_{aq} + (g_{aqjj} - \frac{1}{2} g_{ajjq}) \bar{\kappa}_{ai} \\
    &+ 2 g_{ajqi} \bar{\kappa}_{aj} - \frac{1}{2} g_{aqij} \bar{\kappa}_{aj} 
- \frac{1}{2} g_{aiqj} \bar{\kappa}_{aj}
\end{split}
\end{align}
These terms scale as $O(N^4)$ once expressed in terms of Cholesky vectors, in terms of which
\begin{align}
    \mathcal{F}_{aq}^{\bar{\kappa}} &= \frac{1}{2} \bar{\kappa}_{ai} h_{qi} + \gamma^J D_{qa}^J - \frac{1}{2} L_{qj}^J D_{ja}^J \\
\begin{split}
    \mathcal{F}_{iq}^{\bar{\kappa}} &= \frac{1}{2} \bar{\kappa}_{ai} h_{aq} + E_{qi}^J \gamma^J \\
    &- \frac{1}{2} M_{aq} \bar{\kappa}_{ai} + 2 \delta^J L_{qi}^J - \frac{1}{2} E_{qj}^J L_{ij}^J \\
    &- \frac{1}{2} E_{ij}^J L_{qj}^J,
\end{split}
\end{align}
where 
\begin{align}
    D_{qa}^J &= L_{qi}^J \bar{\kappa}_{ai} \\
    \gamma^J &= L_{jj}^J \\
    E_{qi}^J &= L_{aq}^J \bar{\kappa}_{ai} \\
    M_{aq} &= L_{aj}^J L_{jq}^J \\
    \delta^J &= L_{ai}^J \bar{\kappa}_{ai}.
\end{align}
We have used the Einstein's implicit summation in all these expressions.

\bibliography{bibliography.bib}

\end{document}